\documentclass[a4paper]{article}

\usepackage{bm}  
\usepackage{amsmath}  
\usepackage[dvipdfmx]{graphicx}  

\begin{document}

\title{Phase Separations of Strongly Coupled Fine Particles\\ and Fine Particle Mixtures in Plasmas\footnote{submitted to Conrib. Plasma Phys.}}

\author{Hiroo Totsuji\footnote{e-mail: totsuji-09@t.okadai.jp}\,\\ Graduate School of Natural Science and Technology, Okayama University\footnote{Professor Emeritus}\ ,\\ Japan}

\date{} 

\maketitle 

\abstract{
Phase separations in strongly coupled fine particles in plasmas
are discussed and two-component mixtures are simulated
by molecular dynamics
with the background plasma being treated as continuum.
The system size of laboratory experiments is assumed
and
separations into phases with a common electron density of the background plasma are analyzed.
Since the charge on fine particles increases approximately in proportion to the size,
we expect
the larger component with stronger coupling condensates
from the mixture.
Results expressed in terms of strengths of Coulomb coupling and screening of the larger component
seem to be mostly similar to the one-component case,
at least in cases where
the ratio of fine particle sizes is 2
and the mixing ratio is in the range from 0.25 to 0.75. 
}

\section{Introduction}

When fine particles of radii around microns are placed in classical plasmas
(either fully or partially ionized),
they are negatively charged
to a few times $-(10^2-10^3)e$ ($e$ being the elementary charge).
Though mutual interactions are screened by surrounding electron-ion plasma,
these fine particles form strongly coupled Coulomb-like systems\cite{SM02,FIKKM05,MI09,HT20a}.
The purpose of this article is to discuss possible phase separations of fine particle mixtures
in the domain of strong coupling\cite{HT08a,HT11,HT20,HT23}.

The screening effect of the inert component of the system (background)
defines the difference between one-component plasma (OCP) or ionic mixtures in uniform background
and
the Yukawa system or Yukawa mixtures in ambient plasmas.
In the former,
the charge neutrality of the whole system
is satisfied by the existence of uniform background.
In the latter,
it is satisfied
by the small charge imbalance of electron and ion densities in ambient plasmas
resulting from absorption of electrons onto fine particles. 
(In numerical simulations of uniform systems,
one can forget the charge neutrality of the whole system
in the dynamics of particles 
by adopting the periodic boundary conditions.)

For systems with the background of densities kept uniform,
there have been extensive analyses both theoretical and of simulations\cite{BH80,RKG88etc}.
Here we consider the cases
where
the uniformity of the background cannot be assumed.
In the case of OCP or ionic mixtures,
the background is implicitly assumed to be degenerate electrons
and
the total pressure of the system is mainly determined by electrons.
The large pressure of degenerate electrons 
allows us to consider ions to be immersed in uniform background charge.

For coexisting two phases,
the pressure is common.
When the pressure is mainly determined by the background,
we have the same density of background plasma in two phases
and
the separation of ion mixtures or Yukawa mixtures 
in the uniform background.
In the case of ionic mixtures in uniform background,
two phases have the same ion charge density.
Since the interaction energy and the pressure of strongly coupled OCP 
is approximately proportional to the parameter
$(4\pi n_i/3)^{1/3} (Qe)^2 /4 \pi \epsilon_0 k_B T_i \propto (Qn_i)^2 /n_i^{5/3}$
($n_i$, $Qe$, and $T_i$ being the number density, the charge, and the temperature of ions),
two phases with different charge number do not coexist with common $Qn_i$. 

For Yukawa mixtures with the common charge imbalance in the background plasma,
the situation is similar to the case of ionic mixture with the uniform (common)  background:
The pressure of strongly coupled Yukawa particles is also a monotonic function of the density.
The charge imbalance, however, can be different in two phases
and we have a possibility of coexisting two phases of Yukawa particles.
We can have either different densities of whole background plasma\cite{HT08a}
or different ion densities with a common electron density\cite{HT11,HT20,HT23}.

In the case of OCP or ionic mixtures,
the pressure is mainly determined by the background of degenerate electrons.
In the case of fine particles, on the other hand,
the magnitude of the pressure of fine particles 
can be comparable with the pressure of background classical plasma
when fine particles are very strongly coupled:
Extremely strong coupling can be realized due to large magnitude of fine particle charge.

In strongly coupled Yukawa systems with the background of classical plasma,
the difference in the magnitudes of pressures of background plasma and Yukawa particles 
can be comparable with the opposite signs.
Therefore,
we have a possibility of coexisting two phases with different densities
of both background plasma and Yukawa particles.
Based on an interpolation formula for thermodynamic quantities obtained by numerical simulations,
we obtain the conditions for coexisting phases with different densities
in both the background plasma and Yukawa particles\cite{HT08a}.
We note that,
due to large difference in the density of background plasma and Yukawa particles,
the Yukawa particles need to be in the state of extremely strong coupling.
An example of the phase diagram is shown in Fig.1(Left).

When the Coulomb coupling between fine particles is not very strong,
the pressure of the system is determined by the background plasma.
We can then regard fine particles 
(accompanying increased part of ion density which neutralize the negative charges on fine particles)
as a kind of solute in the solvent of background plasma 
and obtain phase diagrams\cite{HT11,HT20}.
An example is shown in Fig.1(Right).

For Yukawa systems,
we may expect, with the increase of coupling,
to have first the phase separation in uniform electron density
and then the phase separation into phases with different electron densities
in the denser phase \cite{HT23}.

In these investigations,
the system size has not been considered seriously:
sufficiently large system size is implicitly assumed.
This point has been taken into account and
it has been shown that,
when the system size is small (as in laboratories),
we have phase separations with a common electron density\cite{HT23}.
The results are also compared with those of numerical simulations
and,
though parameters are not exactly the same,
theoretical results have been supported.
In this article,
we extend the result to the case of fine particle mixtures.

The similarities and differences of OCP (or ionic mixtures) 
and Yukawa system (or mixtures) in classical plasmas 
are summarized in Table 1.


\begin{table}
\begin{center}
\caption{OCP (ionic mixture) vs. fine particles (mixtures) in plasmas}
\begin{tabular}{l|c|c|c}\hline
\multicolumn{1}{l|}{typical system} & \multicolumn{1}{c|}{OCP, ionic mixture} 
& \multicolumn{2}{c}{ fine particles (dusts) in classical plasmas} \\ \hline
\multicolumn{1}{l|}{charge on ion or fine particle ($e$; elementary charge)} 
& a few $e$ & \multicolumn{2}{c}{${\rm a\ few\ times\ }-(10^2 - 10^3) e$}  \\ \hline
background & degenerate electrons 
& \multicolumn{2}{c}{fully or partially ionized classical plasma} \\ \hline
Coulomb coupling & weak or strong  & weak or strong & very strong  \\ \hline
main contribution to pressure of the system & degenerate electrons & plasma 
& plasma and fine particles  \\ \hline
\end{tabular}
\end{center}
\end{table}

Components of our system are fine particles 
and the weakly ionized plasma 
including neutral gas atoms, electrons, 
and ions.
We assume typical values for densities and temperatures
corresponding to usual fine particle (dusty) plasma experiments
as listed in Table 2.
We take Ar as gas species in order to specify some species-dependent values of parameters
and
expect similar discussions hold also for other inert gases.

\begin{table}[h]
\caption{Typical parameters:
$A_{n,e,i,p}\equiv A$ for neutral gas, electrons, ions, and fine particles,
normalized as $p_n[10{\rm Pa}] \equiv p_n/10\ {\rm Pa}$.}
\begin{center}
\begin{tabular}{ll} \hline
neutral gas atoms & \\
\hspace{0.5cm} species & Ar (or other inert gas)\\
\hspace{0.5cm} pressure, temperature & $p_n \sim (10-10^2)\ {\rm Pa},\ T_n \sim 300\ {\rm K}$ \\
\hspace{0.5cm} density 
& $n_n \approx 2.4\cdot 10^{15}\ p_n {\rm [10Pa]}\ T_n {\rm [300K]} \ {\rm cm^{-3}}$ \\ \hline
electrons & \\
\hspace{0.5cm} density, mean distance &
 $n_e \sim (10^8-10^9)\ {\rm cm^{-3}},\ a_e=(3/4\pi n_e)^{1/3} \sim 10^{-3}\ {\rm cm}$ \\
\hspace{0.5cm} temperature & $k_B T_e \sim (1-3)\ {\rm eV}$ \\
ions $({\rm Ar^+})$ & \\
\hspace{0.5cm} density, mean distance 
& $n_i \sim (10^8-10^9)\ {\rm cm^{-3}},\ a_i=(3/4\pi n_e)^{1/3} \sim 10^{-3}\ {\rm cm}$\\
\hspace{0.5cm} temperature & $T_i \sim 300\ {\rm K}$ \\
fine particles & \\
\hspace{0.5cm} size (radius) & $r_p \sim 1\ {\rm \mu m}$ \\
\hspace{0.5cm} density, mean distance 
& $n_p \sim (10^4-10^5)\ {\rm cm^{-3}},\ a_p=(3/4\pi n_p)^{1/3} \sim 10^{-2}\ {\rm cm}$\\
\hspace{0.5cm} temperature & $T_p \sim 300\ {\rm K}$ \\
\hspace{0.5cm} charge & $-Qe,\ Q \sim 3.5 \cdot 10^2\ r_p[{\rm \mu m}]\ k_B T_e[{\rm eV}]$ \\
\hspace{0.5cm} charge density & $|-Qe n_p| \ll en_e \sim en_i$ \\ \hline
Debye lengths & \\
\hspace{0.5cm} ions & $\lambda_i 
\approx 1.2 \cdot 10^{-2}\ (T_i[300{\rm K}])^{1/2}\ (n_i[10^8{\rm cm^{-3}}])^{-1/2}\ {\rm cm}$ \\
\hspace{0.5cm} electrons & $\lambda_e 
\approx 7.4 \cdot 10^{-2}\ (k_BT_e[{\rm eV}])^{1/2}\ (n_e[10^8{\rm cm^{-3}}])^{-1/2}\ {\rm cm}$ \\
\hline
\end{tabular}
\end{center}
\end{table}

The number $Q$ of charges on fine particles $-Qe$ is approximately given by\cite{FIKKM05}
\begin{align}
Q &\sim 0.5 \frac{k_B T_e} {e^2/4 \pi \varepsilon_0 r_p} 
= 3.5 \cdot 10^2\ r_p[{\rm \mu m}]\ k_B T_e[{\rm eV]} \gg 1.
\end{align}
Though $Q \gg 1$,
we still have $n_e \sim n_i \gg Q n_p$.
These charges on particles are screened by surrounding ion-electron plasma
and
we have a system of particles interacting via the screened Coulomb or the Yukawa potential
\begin{align}\label{Yukawa}
&\frac{(Qe)^2}{4 \pi \varepsilon_0 r}\exp\left(-\frac{r}{\lambda} \right),
\end{align}
where $\lambda$ is the Debye screening length (inverse of the Debye wave number) 
defined by (typical values shown in Table 2)
\begin{align}
\frac{1}{\lambda^2}=\frac{1}{\lambda_e^2}+\frac{1}{\lambda_i^2}
\sim \frac{1}{\lambda_i^2},
\ \frac{1}{\lambda_{e,i}^2}=\frac{e^2n_{e,i}}{\varepsilon_0 k_B T_{e,i}}.
\end{align}
In our system,
we have two characteristic parameters, $\Gamma$ and $\xi$, defined respectively by
\begin{align}
\Gamma \equiv \frac{(Qe)^2}{4 \pi \varepsilon_0 a_p k_B T_p}
\ {\rm and}\ \ \xi \equiv \frac{a_p}{\lambda},
\end{align}
where the mean distance $a_p$ is defined by the fine particle density $n_p$ as
\begin{align}
a_p = \left(\frac{3}{4\pi n_p}\right)^{1/3}.
\end{align}

\begin{figure}
\begin{center}
\hspace*{-2cm}
\includegraphics[width=10.0cm]{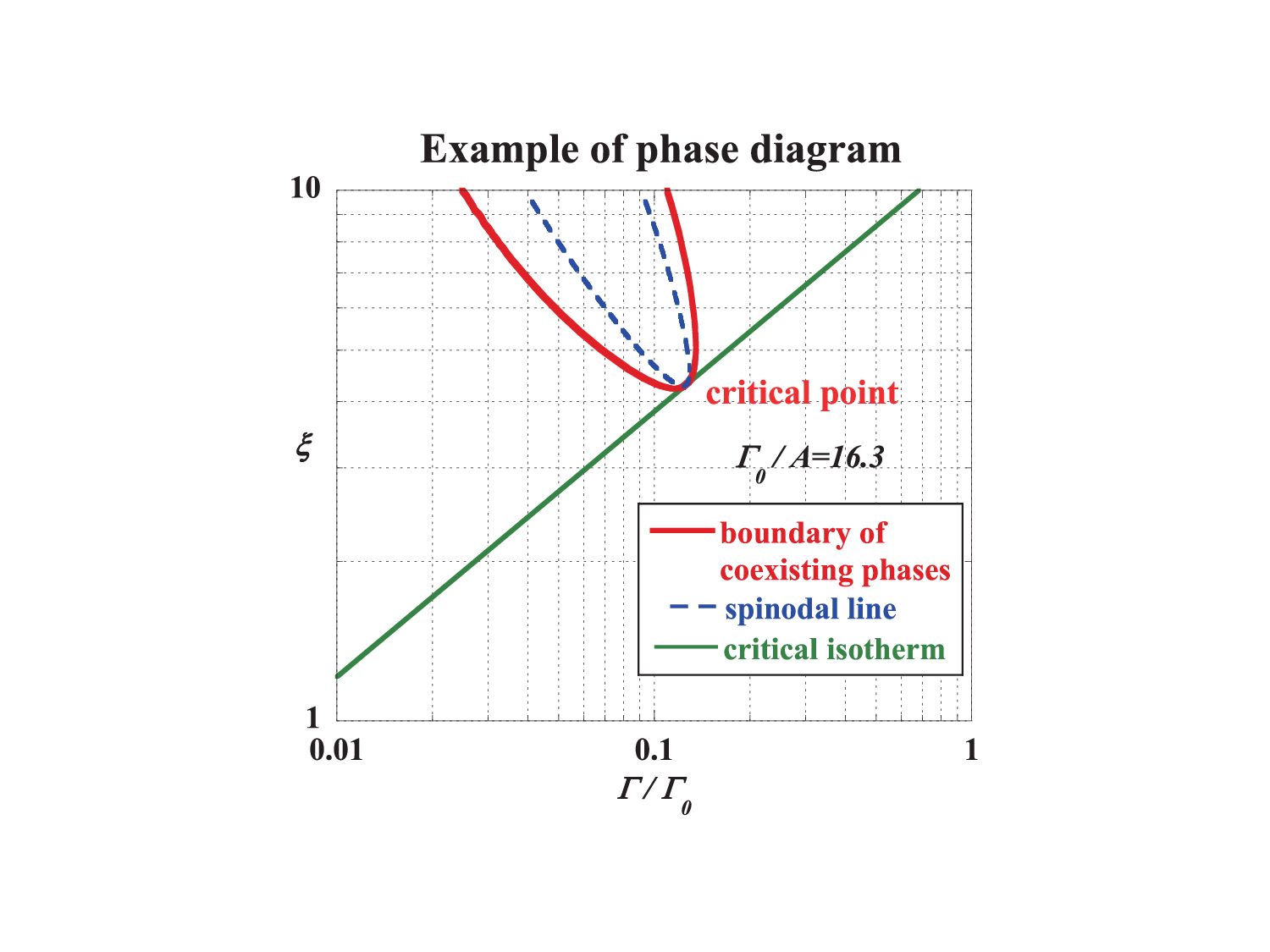}
\includegraphics[width=6.0cm]{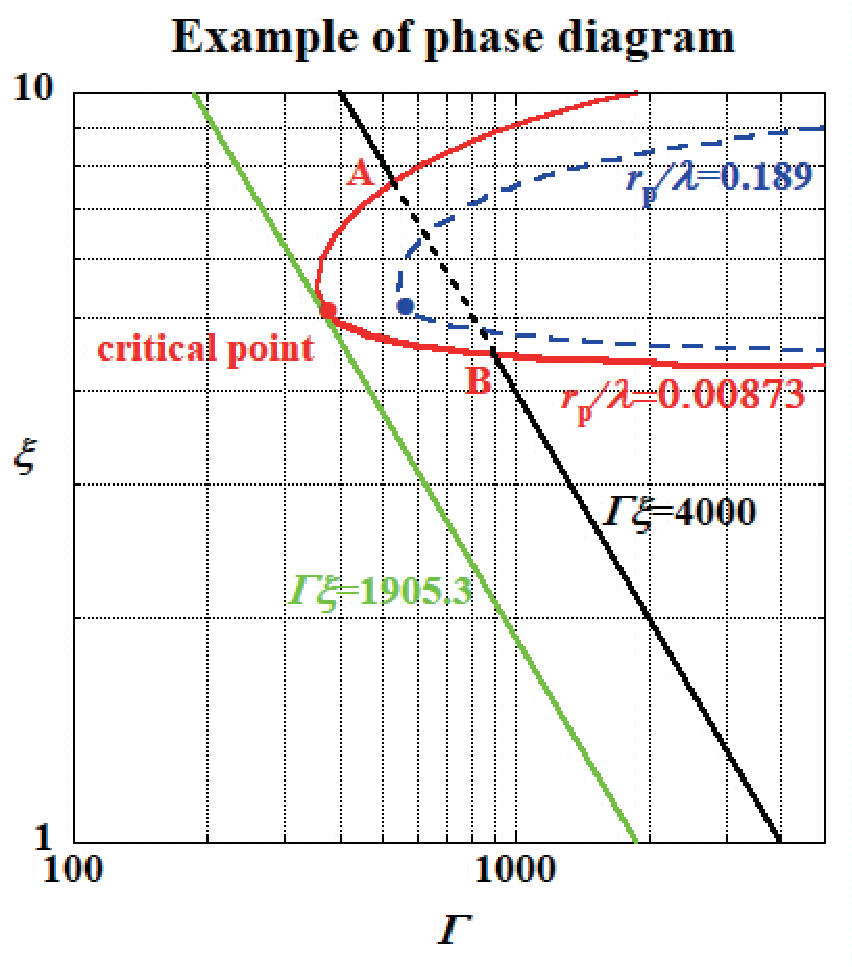}
\caption{{\bf (Left)} Separation into phases with different plasma densities\cite{HT08a}
($A \sim 10^5 \gg 1$ coming from the ratio of ideal gas pressure of electrons to that of fine particles).
{\bf (Right)} Separation into phases with a common electron density\cite{HT20}.}
\label{Fig.1}
\end{center}
\end{figure}


\label{sec1}
\section{Behavior of Components}

There are four components in our system:
Neutral gas atoms (density $n_n$, temperature $T_n$),
the electron-ion plasma
($n_e \sim n_i$, $T_e$, $T_i$),
and fine particles ($n_p$, $T_p$)
with typical values of parameters in Table 2.
In order to have coexisting phases, the system has to be large enough in comparison with the thickness of boundary layers between two phases.  The latter may be evaluated by the mean free paths of collisions.

\subsection{Collision Frequencies and Temperatures}

Numerical values related to collisions between neutral gas atoms and electrons and ions 
are listed in Table 3
and
those related to Coulomb collisions are listed in Table 4.
The superscripts $(E)$ indicate those for the energy transfer.
We observe the assumption for the temperatures are consistent with these values.

\begin{table}[h]
\caption{Collisions between neutral atoms, electrons and ions}
\begin{center}
\begin{tabular}{ll} \hline
electron-neutral & \\
\hspace{0.5cm} cross section & $\sigma_{en} \sim 2 \cdot 10^{-16}\ {\rm cm^2}$ \\
\hspace{0.5cm} collision frequency & $\displaystyle  \nu_{en}=n_n \sigma_{en} v_{th,e} 
\sim 2 \cdot 10^7\ \frac{p_n[10{\rm Pa}](k_B T_e[{\rm eV}])^{1/2}}
{T_n[300{\rm K}]}\ {\rm s^{-1}}$ \\
\hspace{0.5cm} 
& $\displaystyle  \nu_{en}^{(E)}=\nu_{en} \frac{m_e}{m_n} 
\sim 3 \cdot 10^2\ \frac{p_n[10{\rm Pa}](k_B T_e[{\rm eV}])^{1/2}}
{T_n[300{\rm K}]}\ {\rm s^{-1}}$ \\
\hspace{0.5cm} mean free path 
& $\displaystyle \ell_{en}=\frac{1}{n_n \sigma_{en}} 
\sim 2\ \frac{T_n[300{\rm K}]}{p_n[10{\rm Pa}]}\ {\rm cm}$ \\
ion-neutral & \\
\hspace{0.5cm} cross section & $\sigma_{in} \sim 1 \cdot 10^{-14}\ {\rm cm^2}$ \\
\hspace{0.5cm} collision frequency 
& $\displaystyle  \nu_{in}=n_n \sigma_{in} v_{th,i} 
\sim 6 \cdot 10^5\ \frac{p_n[10{\rm Pa}](T_i[300{\rm K}])^{1/2}}
{T_n[300{\rm K}]}\ \ {\rm s^{-1}}$ \\ 
\hspace{0.5cm} mean free path 
& $\displaystyle \ell_{in}=\frac{1}{n_n \sigma_{in}} 
\sim 5 \cdot 10^{-2}\ \frac{T_n[300{\rm K}]}{p_n[10{\rm Pa}]}\ {\rm cm}$ \\
neutral-neutral & \\
\hspace{0.5cm} cross section & $\sigma_{nn} \sim 1 \cdot 10^{-15}\ {\rm cm^2}$ \\
\hspace{0.5cm} collision frequency 
& $\displaystyle  \nu_{nn}=n_n \sigma_{nn} v_{th,n} 
\sim 7 \cdot 10^4\ \frac{p_n[10{\rm Pa}]}{(T_n[300{\rm K}])^{1/2}}\ {\rm s^{-1}}$ \\
\hspace{0.5cm} mean free path 
& $\displaystyle \frac{1}{n_n \sigma_{nn}} 
\sim 4 \cdot 10^{-1}\ \frac{T_n[300{\rm K}]}{p_n[10{\rm Pa}]}\ {\rm cm}$ \\
\hline
\end{tabular}
\end{center}
\end{table}

\begin{table}[h]
\caption{Coulomb collisions}
\begin{center}
\begin{tabular}{ll} \hline
electron-electron & \\
\hspace{0.5cm} cross section & $\sigma_{ee} 
\sim 2.3 \cdot 10^{-12} (T_e{\rm[eV]})^{-2}\ {\rm cm^2}$ \\
\hspace{0.5cm} collision frequency 
& $\displaystyle  \nu_{ee}=n_e \sigma_{ee} v_{th,e} 
\sim 9.8 \cdot 10^3\ \frac{n_e[10^8{\rm cm^{-3}}]}{(T_e[{\rm eV}])^{3/2}}
\ {\rm s^{-1}}$ \\
\hspace{0.5cm} mean free path 
& $\displaystyle \frac{1}{n_e \sigma_{ee}} 
\sim 4.3 \cdot 10^3 \frac{(T_e[{\rm eV}])^2}{n_e[10^8{\rm cm^{-3}}]}
\ {\rm cm}$ \\
ion-ion & \\
\hspace{0.5cm} cross section & $\displaystyle \sigma_{ii} 
\sim 2.3 \cdot 10^{-9} (T_i{\rm [300K]})^{-2} {\rm cm^2}$ \\
\hspace{0.5cm} collision frequency & $\displaystyle  \nu_{ii}=n_i \sigma_{ii} v_{th,i} 
\sim 5.8 \cdot 10^3\ \frac{n_i[10^8{\rm cm^{-3}}]}{(T_i[300{\rm K}])^{3/2}}
\ {\rm s^{-1}}$ \\
\hspace{0.5cm} mean free path 
& $\displaystyle \frac{1}{n_i \sigma_{ii}} 
\sim 4.3 \ \frac{(T_i[300{\rm K}])^2}{n_i[10^8{\rm cm^{-3}}]}\ {\rm cm}$ \\
electron-ion & \\
\hspace{0.5cm} cross section & $\displaystyle \sigma_{ei} \sim \sigma_{ee}$ \\
\hspace{0.5cm} collision frequency & $\displaystyle  \nu_{ei} \sim \nu_{ee}$ \\
\hspace{0.5cm}  & $\displaystyle  \nu_{ei}^{(E)}=\nu_{ei} \frac{m_e}{m_i}
\sim 1.4 \cdot 10^{-1}\ \frac{n_i[10^8{\rm cm^{-3}}]}{(T_e[{\rm eV}])^{3/2}}
\ {\rm s^{-1}}$ \\
\hspace{0.5cm} mean free path
& $\displaystyle \frac{1}{n_i \sigma_{ei}} \sim \frac{1}{n_e \sigma_{ee}}$ \\
\hline
\end{tabular}
\end{center}
\end{table}

\subsection{Collision Mean Free Path and System Size}

We note that
the electron-neutral collision mean path is larger than other electron mean paths
(the ones for collision with ions and electrons)
and estimated as\cite{LL05}
\begin{align}
&\frac{1}{n_n \sigma_{en}}
\sim 2\ \frac{T_n[300{\rm K}]}{p_n[10{\rm Pa}]}\ {\rm cm}
\left( \gg 
\frac{1}{n_n \sigma_{in}} 
\sim 5 \cdot 10^{-2}\ \frac{T_n[300{\rm K}]}{p_n[10{\rm Pa}]}\ {\rm cm}\right).\label{en-in}
\end{align}
Therefore,
the electron density may have no structures
at least
with the scale less than this mean free path.
\label{sec2-1}
\subsection{Application of Drift-Diffusion Equations}

We can discuss the charge neutrality and the potential structure in fine particle clouds in plasmas
on the basis of drift-diffusion equations\cite{HT16ab,HT17ab,HT18}.

\noindent
{\bf Characteristic scales of length}

We have three kinds of scale lengths in our system:
the system size $R_0$,
mean free paths for electrons $\ell_e$, ions $ \ell_i$, and fine particles $\ell_p$, 
and
mean distances between electrons $a_e=(3/4 \pi n_e)^{1/3}$, ions $a_i=(3/4 \pi n_i)^{1/3}$, 
and fine particles $a_p=(3/4\pi n_p)^{1/3}$,
where $a_e \sim a_i \sim 10^{-3}\ {\rm cm}$ and $a_p \sim 10^{-2}\ {\rm cm}$.

We can describe the system in terms of densities $n_{e, i, p}$
in a macroscopic scale $L$ satisfying $R_0 \gg L \gg a_p \gg a_e\sim a_i$
and,
taking $L$ so that $R_0 \gg L \gg \ell$ ($\ell$ being the collision mean free path),
we can apply drift-diffusion equations 
with appropriate diffusion coefficients $D$ and mobilities $\mu$.

In our system,
all components (electrons, ions, and fine particles) collide mainly with neutral gas atoms
 (cf. Table 3) and
\begin{align}
\ell_{en} \gg \ell_{in}, \ell_{pn},
\end{align}
where $\ell_{en}, \ell_{in}$, and $\ell_{pn}$ are 
electron-neutral, ion-neutral, and particle-neutral collision mean free paths, respectively.
The value of $\ell_{en}$ depends on the neutral gas pressure $p_n$ and temperature $T_n$ as (\ref{en-in})
and typically, $\ell_{en}\sim 0.2\ {\rm cm}\ (p_n=10^2\ {\rm Pa})$
and $\ell_{en}\sim 2\ {\rm cm}\ (p_n=10\ {\rm Pa})$.
We therefore have two cases:
\begin{align}
R_0 \gg \ell_{en}\ (\gg \ell_{in}, \ell_{pn}), \label{case1}\\
R_0 \sim \ell_{en}\ (\gg \ell_{in}, \ell_{pn}).\label{case2}
\end{align}
When (\ref{case1}), the drift-diffusion equation is applicable to all components,
and
when (\ref{case2}),  only to ions and fine particles.


When (\ref{case1}), we can take the length $L$ so that $R_0 \gg L \gg \ell_{en}$
and have drift-diffusion equations for densities:
\begin{align*}
&{\bm \nabla}\cdot (-D_e {\bm \nabla}n_e - \mu_e n_e {\bm E}) = \frac{\delta n_e}{\delta t}
= c_g n_e - c_p n_p,\\
&{\bm \nabla}\cdot (-D_i {\bm \nabla}n_i + \mu_i n_i {\bm E}) = \frac{\delta n_i}{\delta t}
= c_g n_e - c_p n_p,\\
&{\bm \nabla}\cdot (-D_p {\bm \nabla}n_p - \mu_p n_p {\bm E}) = 0.
\end{align*}
Here $D_{e,i,p}$ and $\mu_{e,i,p}$ are 
diffusion coefficients and mobilities of electrons, ions, and fine particles, respectively 
(we assume Einstein relations
$D_{e,i}=e \mu_e/k_B T_{e,i}$ and $D_p=Qe \mu_p/k_B T_p$).
On right-hand sides,
the generation of electrons and ions, $\delta n_e/\delta t =\delta n_i/\delta t$,
are given by the difference in the impact ionization of neutral atoms by electrons with the rate $c_g$ 
and the absorption onto fine particles with the rate $c_p$.

From these equations,
we have
\begin{align}
&\frac{|en_i-en_e-Qen_p|}{en_e} 
\sim \frac{\lambda_e^2}{R_a^2} \frac{1}{Q^2n_p/n_i+1} \ll 1,
\label{qcn}
\end{align}
where 
$R^2_a=D_a/c_g$, $D_a=(\mu_eD_i+\mu_iD_e)/(\mu_e+\mu_i)$ being the ambipolar diffusion coefficient.
Since $R_a \sim 1\ {\rm cm}$ and $\lambda_e \leq 10^{-1}\ {\rm cm}$,
we have the quasi-charge-neutrality
enhanced by the contribution of $Q^2n_p \gg n_i$\cite{HT17ab, HT18}.

The change of the average electrostatic potential $\overline{\Psi}$ in the system is estimated as
\begin{align*}
&|\overline{\Psi(R_0)}-\overline{\Psi(0)}| \sim \frac{|en_i-en_e-Qen_p|}{\varepsilon_0} R_0^2 
\sim \frac{R_0^2}{R_a^2}\frac{\lambda_e^2}{\varepsilon_0}
\frac{1}{Q^2n_p/n_i+1} en_e. 
\end{align*}
For electrons, we have
\begin{align}
&\frac {e|\overline{\Psi(R_0)}-\overline{\Psi(0)}|}{k_B T_e} 
\sim \frac{R_0^2}{R_a^2}\frac{1}{Q^2n_p/n_i+1}.\label{qcn2}
\end{align}
Therefore,
when $Q^2n_p/n_i \gg 1$ and $R_0 \sim R_a$,
we may assume the electron density is almost constant in and out of the fine particle cloud.
For ions, on the other hand,
\begin{align}
&\frac {e|\overline{\Psi(R_0)}-\overline{\Psi(0)}|}{k_B T_i} 
\sim \frac{R_0^2}{R_a^2}\frac{T_e/T_i}{Q^2n_p/n_i+1}
\end{align}
indicates that
the ion density change may be appreciable even when $Q^2n_p/n_i \gg 1$.
These results are consistent with numerical solutions of drift-diffusion equations\cite{HT17ab, HT18}.
Separating the ion density $n_i$ into $n_i-n_i^{(0)}$ and $n_i^{(0)}=n_e$,
we have
\begin{align}
&n_i-n_e=(n_i^{(0)}-n_e)+(n_i-n_i^{(0)})=n_i-n_i^{(0)} \sim Q n_p
\end{align} 
or,
ions and fine particle are added into the system so that
the fine particle charge is compensated by the increase of ions,
supporting our treatment of them as solute in the solvent of electron-ion plasma.

As far as the size of fine particle cloud $R_0$ is comparable with $R_a$,
the enhanced charge neutrality (\ref{qcn}) and (\ref{qcn2}) 
thus justifies the phase equilibrium with a common electron density.
On the other hand,
when $R_0 \gg R_a$,
the electron density changes appreciably and
we have to consider the phase equilibrium 
with different densities of all components\cite{HT08a}.


When (\ref{case2}),
we may treat the electron distribution $n_e$ as an ideal gas
in the average electrostatic potential $\overline{\Psi({\bm r})}$
or
\begin{align*}
&n_e \sim n_e^{(0)} \exp\left(\frac{e\overline{\Psi({\bm r})}}{k_B T_e}\right),
\end{align*}
$n_e^{(0)}$ being position independent.
Then we have
\begin{align*}
&{\bm \nabla}\cdot \left({\bm \nabla}(-e n_e) - \frac{1}{\lambda_e^2}\varepsilon_0{\bm E}\right) \sim 0.
\end{align*}
Using this relation,
we have the same conclusion as in the case of (\ref{case1}):
We have the quasi-charge-neutrality and the flatness of the electron distribution
also in this case.

\label{sec2-2}
\section{Numerical Simulations}

\subsection{Helmholtz free energy}

We define the local average $\overline{A({\bm r}, t)}$ 
of a microscopic quantity $A({\bm r}, t)$
over a volume $a_0^3$ centered at ${\bm r}$ by
\begin{align*}
&\overline{A({\bm r}, t)} \equiv \frac{1}{a_0^3}\int_{a_0^3} d{\bm r}' A({\bm r}', t),
\end{align*}
taking $a_0$ so that $R_0 \gg a_0 \gg a_p (\gg a_{e,i})$:
For example,
densities are defined by microscopic densities $\hat{\rho}_{e, i, p}({\bm r},t)$ as
$n_{e, i, p}({\bm r},t) \equiv  \overline{\rho_{e, i, p}({\bm r},t)}
=\overline{\sum_j \delta [{\bm r}-{\bm r}_j(t)]}$.

For the system of electrons, ions, and fine particles
where the average densities change with characteristic length $R_0$ 
satisfying the charge neutrality in the scale of $a_0$,
we have obtained\cite{HT15} 
the effective interaction for fine particles $U_{ex}$
expressed by the Helmholtz free energy of electron-ion system
for the given configuration of fine particles $\{{\bm r}_i\}_{i=1, ..,N}$:
\begin{align}
U_{ex}
=
F_{id,0} 
&+
{1 \over 2} \left[ \int \int d{\bm r}d{\bm r}' u({\bm r},{\bm r}')
\left[\hat{\rho}_p({\bm r})-(-Qe)n_p({\bm r})\right]
\left[\hat{\rho}_p({\bm r}')-(-Qe)n_p({\bm r}')\right]
- (s. i.)\right] \nonumber \\
&-{1 \over 2} \sum_{i=1}^N {(Qe)^2 k_D({\bm r}_i) \over 4 \pi \varepsilon_0},
\label{Helmholtz}
\end{align}
where
\begin{align}
&F_{id,0}=k_B T_e \int d{\bm r} 
{n_e}({\bm r}) \left[\ln [{n_e}({\bm r}) \Lambda_e^3] -1 \right]
+
k_B T_i \int d{\bm r} 
{n_i}({\bm r}) \left[\ln {n_i}({\bm r}) \Lambda_i^3] -1 \right]
\label{Helmholtz-0}
\end{align}
and
\begin{align*}
u({\bm r},{\bm r}') 
={\exp(-k_D^+ |{\bm r}-{\bm r}'|) \over 4 \pi \varepsilon_0 |{\bm r}-{\bm r}'|}
\end{align*}
with the Debye wave number $k_D=1/\lambda$ at $({\bm r}+{\bm r}')/2$,
$k_D^+ = k_D [({\bm r}+{\bm r}')/2]$.
(We assume the change in the ion density is small
in the scale of mean fine-particle distance.)
The trivial extension into the case of multi-component fine particles
will not be shown here and in what follows.

The term $F_{id,0}$ is 
the ideal gas Helmholtz free energy of average electron and ion distributions
($\Lambda_{e,i}$ being thermal de Broglie wavelengths of electrons and ions).
The second term of $U_{ex}$ is the effective interaction between fine particles.
It is to be noted that
they are interacting via the repulsive Yukawa potential
and, at the same time,
confined by the `shadow' charge density $-(-Qe) n_p({\bm r},t)=Qe n_p({\bm r},t)$
which cancels the average charge density of fine particles
$\overline{\rho_p({\bm r},t)}=(-Qe)n_p({\bm r},t)$:
Since the system as a whole is charge neutral,
fine particles are not simply repelling each other\cite{HT15, HF94a, YR94}.
The term $-(s.i)$ subtracts self-interactions $\sum_i u({\bm r}_i,{\bm r}_i)$ 
formally included in the integral.
The third term is the energy stored in polarization cloud around each fine particle
called polarization potential
which gives the so-called polarization force\cite{HF94b}.

By numerical simulations based on (\ref {Helmholtz}) and (\ref {Helmholtz-0}),
we can include strong coupling effects between fine particles
which are not included in analyses by drift-diffusion equations.
Though the screening effect is essential,
we regard electrons and ions themselves in the background plasma as ideal gases.


\subsection{Equations of motion}

Equations of motion for fine particles at $\{{\bm r}_i\}_{i=1,...N}$
are summarized as\cite{HT23}
\begin{align*}
&{d \over d t} {\bm r}_i (t) =  {\bm v}_i(t),\ \ \ 
m_i {d \over d t} {\bm v}_i (t) =  {\bm F}_i ({\bm r}_i, t)
= -{\partial \over \partial {\bm r_i}} \Phi (\{{\bm r}_j\}, t)
- \left[ {\partial \over \partial {\bm r}} 
[\phi_s ({\bm r},t) +\phi_{pp} ({\bm r},t) + \phi_p ({\bm r},t)]
\right]_{{\bm r}={\bm r}_i},
\end{align*}
where
\begin{align*}
&\Phi (\{{\bm r}_j\}, t) = {Q^2 e^2 \over 4 \pi \varepsilon_0 }\ 
{1 \over 2} \sum_{i \neq j} {1 \over r_{ij}} \exp(- k_D r_{ij}), \\
&\phi_s ({\bm r}, t)
=   - \int d{\bm r}' \frac{Q^2e^2}{4\pi \varepsilon_0 |{\bm r}-{\bm r}'|}
\exp(-k_D |{\bm r}-{\bm r}'|){n_p({\bm r}', t)},\\ 
&\phi_{pp}({\bm r}, t) 
= - \frac{1}{2} \frac{(Qe)^2 k_D ({\bm r}, t)}{4 \pi \varepsilon_0},
\end{align*}
and
\begin{align*}
&\phi_p({\bm r}, t) =Q k_B T_i \ln [n_i^0+Q{n_p} ({\bm r}, t)].
\end{align*}
Here, $k_D({\bm r}, t)$ is the local Debye wave number $1/\lambda$
which depends on the average density of ions $n_i({\bm r}, t)$.
Fine particles with charge $-Qe$ are moved 
by mutual Yukawa repulsion ($\Phi$),
effective attraction by `shadow' charges ($\phi_s$),
the polarization force ($\phi_{pp}$), 
and the ion pressure ($\phi_p$):
Due to charge neutrality in fine particle clouds,
the ion density $n_i^{(0)}+Q{n_p} ({\bm r}, t)$ depends on ${\bm r}$
and its gradient exerts (ideal gas) pressure onto fine particles
which are coupled with ions.

We simulate fine particles by molecular dynamics
adopting the periodic boundary conditions with the cubic unit cell
which contains $N=(1 - 2)\cdot 10^3$ independent fine particles.
Ion and fine particle distributions, $n_i({\bm r}, t)$ and $n_p({\bm r}, t)$, 
are treated as continuous functions of space 
and are represented by their values at mesh points
located at centers of small cells.
Values of $\phi_s$, $\phi_{pp}$, and $\phi_p$ 
and their gradients are computed at mesh points
and
corresponding forces at the position of particle 
are given by linear interpolations.
The number of mesh points is $4^3$ in the unit cell.

The results of one-component case is shown in Fig.2\cite{HT23}.
We observe that,
in the results of numerical simulations,
the position of the phase-coexistence line is shifted 
to smaller $\Gamma$ values,
when compared with theoretical prediction\cite{HT20}.
These differences might be attributed to 
the approximate treatment of the finite radius effect 
in the interpolation\cite{HT08a}
and finite size of numerical simulations.
Though the exact agreement is not obtained,
we expect this kind of phase separations occur
in strongly coupled fine particle plasmas
around these values in the $\Gamma \xi$-plane.
\begin{figure}
\begin{center}
\includegraphics[width=8cm]{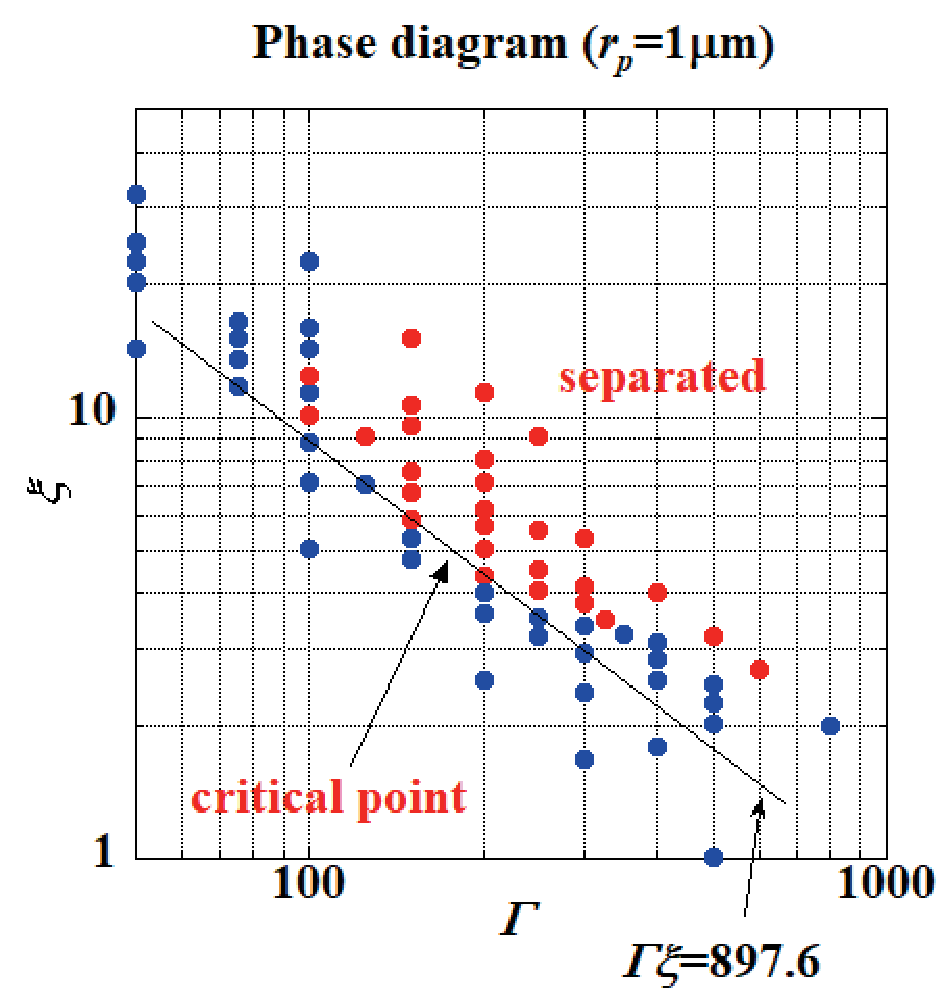}
\caption{Results of numerical simulations 
(phase separation at red points)\cite{HT23}.
The critical point is estimated at $(\Gamma_c=1.80\cdot 10^2, \xi_c=5.0)$ 
on $\Gamma \xi \approx 9.0\cdot 10^2$ line.}
\end{center}
\end{figure}

\subsection{Mixtures of fine particles in plasmas}

We here consider the simplest case
with two kinds of fine particles in the background plasma of electrons and ions.
We assume that
the electron density is uniform and the charge neutrality is satisfied
by increase in the density of ions.

Our system is composed 
of $N_1$ fine particles of radius $r_{p1}$ and $N_2$ fine particles of radius $r_{p2}$
in the volume $V$.
In the background plasma,
the electron density $n_e(=n_i^{(0)})$ is constant,
while the ion density $n_i({\bm r},t)$ includes the uniform part $n_i^{(0)}$
and the $({\bm r},t)$-dependent part $n_i({\bm r},t)-n_i^{(0)}$:
the latter keeps the charge neutrality depending on the density of fine particles
$n_{p1}({\bm r},t)$ and $n_{p2}({\bm r},t)$.
The charges on fine particles, $Q_1$ and $Q_2$, are proportional to the radii
and we assume $r_{p1}=1\ {\rm \mu m}$, $r_{p2}=2\ {\rm \mu m}$
and therefore $Q_2/Q_1=2$.

We have performed numerical simulations for the cases
where neither of components can be regarded as a kind of impurity,
or for the cases with the mixing ratio $x=N_1/(N_1+N_2)=$ 0.25, 0.5, and 0.75.
Typical values of temperatures are taken:
\begin{align*}
k_BT_e=3\ {\rm eV},\ T_i=T_p(=T_{p1}=T_{p2})= 300\ {\rm K}.
\end{align*}
Since $(Q_2/Q_1)^2=4 > 1$,
we expect the separation of phases occurs 
as the condensation of species 2 from the mixture.
For the species 2,
the Coulomb coupling parameter $\Gamma_2$ and the strength of screening $\xi_2$
are given respectively by
\begin{align}
\Gamma_2 = \frac{Q_2^2 e^2}{4 \pi \varepsilon_0 a_2 k_B T_p},
\ \xi_2=\frac{a_2}{\lambda},
\end{align}
where $a_2$ is the mean distance between particles of species 2,
\begin{align}
a_2=\left(\frac{3}{4 \pi n_2}\right)^{1/3}=\left(\frac{3V}{4 \pi N_2}\right)^{1/3}.
\end{align}

In the case of ionic mixtures,
the one-fluid approximation for thermodynamic quantities 
with $Q^2_{eff}=\left(\overline{Q}\right)^{1/3}\overline{Q^{5/3}}$
is known to work when the ratio of charges is small\cite{BH80}.
We have changed 
the effective Coulomb coupling parameter in the range
\begin{align}
100 \leq \Gamma_{eff}= \frac{Q^2_{eff}e^2}{4 \pi \varepsilon_0 a_p k_B T_p}
= \frac{Q^2_{eff}e^2}{4 \pi \varepsilon_0 k_B T_p}\left(\frac{4 \pi (N_1+N_2)}{3 V}\right)^{1/3}
\leq 600
\end{align}
and the strength of screening or the density of electron-ion plasma in the range
\begin{align}
5 \cdot 10^7\ {\rm cm^{-3}} \leq n_e=n_i^{(0)} \leq 5 \cdot 10^8\ {\rm cm^{-3}}.
\end{align}
Values of $\Gamma_2$ and $\Gamma_2 \xi_2$ are respectively given by
\begin{align*}
&\frac{\Gamma_2}{\Gamma_{eff}} =\frac{(1-x)^{1/3}}{[x+(1-x)q]^{1/3}[x+(1-x)q^{5/3}]} q^2
\ \ {\rm with}\ \ q=\frac{Q_2}{Q_1}\\
&\Gamma_2 \xi_2 \sim 2.02 \cdot 10^3\ (n_e\ [10^8\ {\rm cm^{-3}}])^{1/2}.
\end{align*}

Some examples of phase separation between two species of fine particles are shown in Figs.3 and 4.
These changes in distributions are observed
when we increase the value of $\Gamma_2$ along the constant $\Gamma_2 \xi_2$-line
with $n_e=1\cdot 10^8\ [{\rm cm^{-3}}]$ and $x=0.5$.
We clearly observe the appearance of separated phases.
(These two sets of  $(\Gamma_2, \xi_2)$ are neighboring points in the left panel in Fig.7.)
\begin{figure}[h]
\hspace*{0.2cm}
\includegraphics[width=80mm]{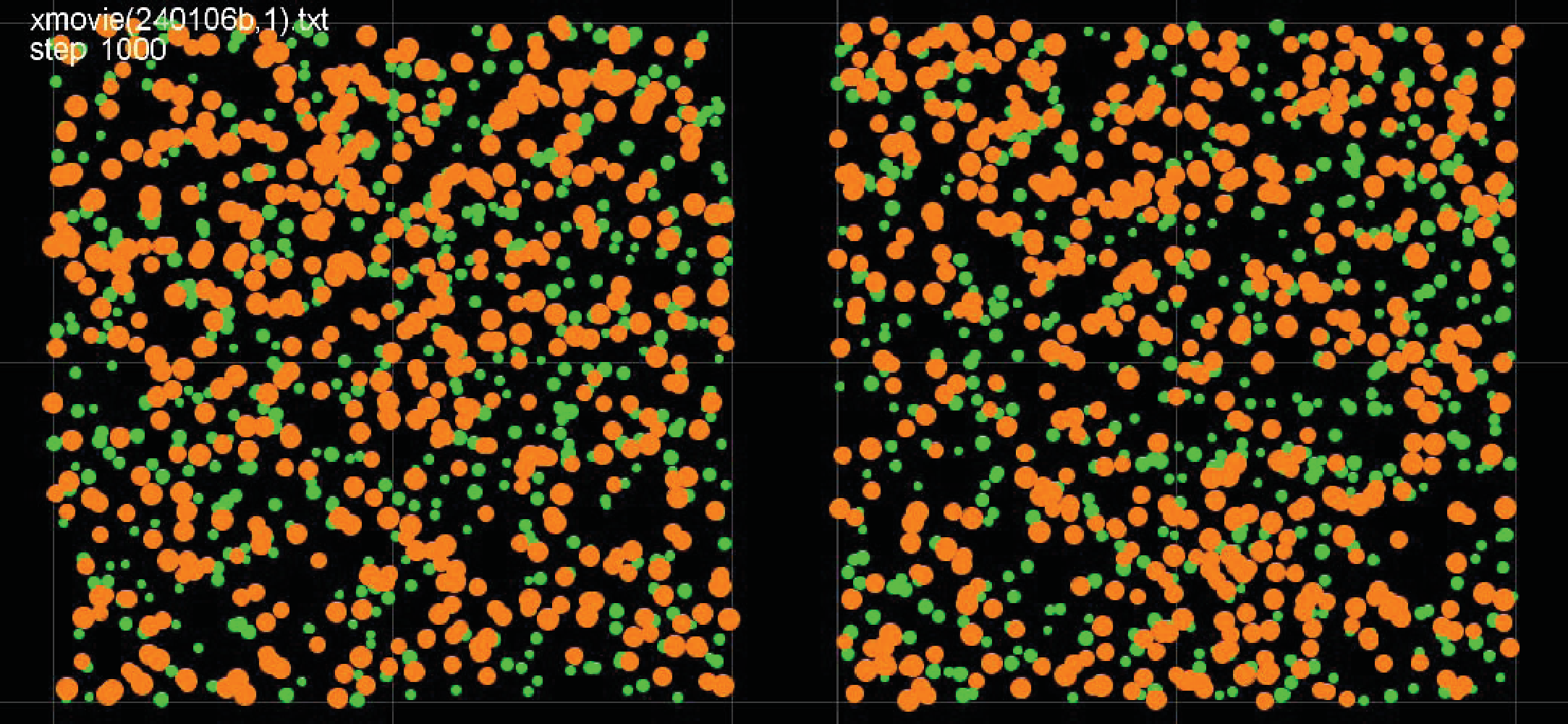}
\hspace*{1cm}
\includegraphics[width=80mm]{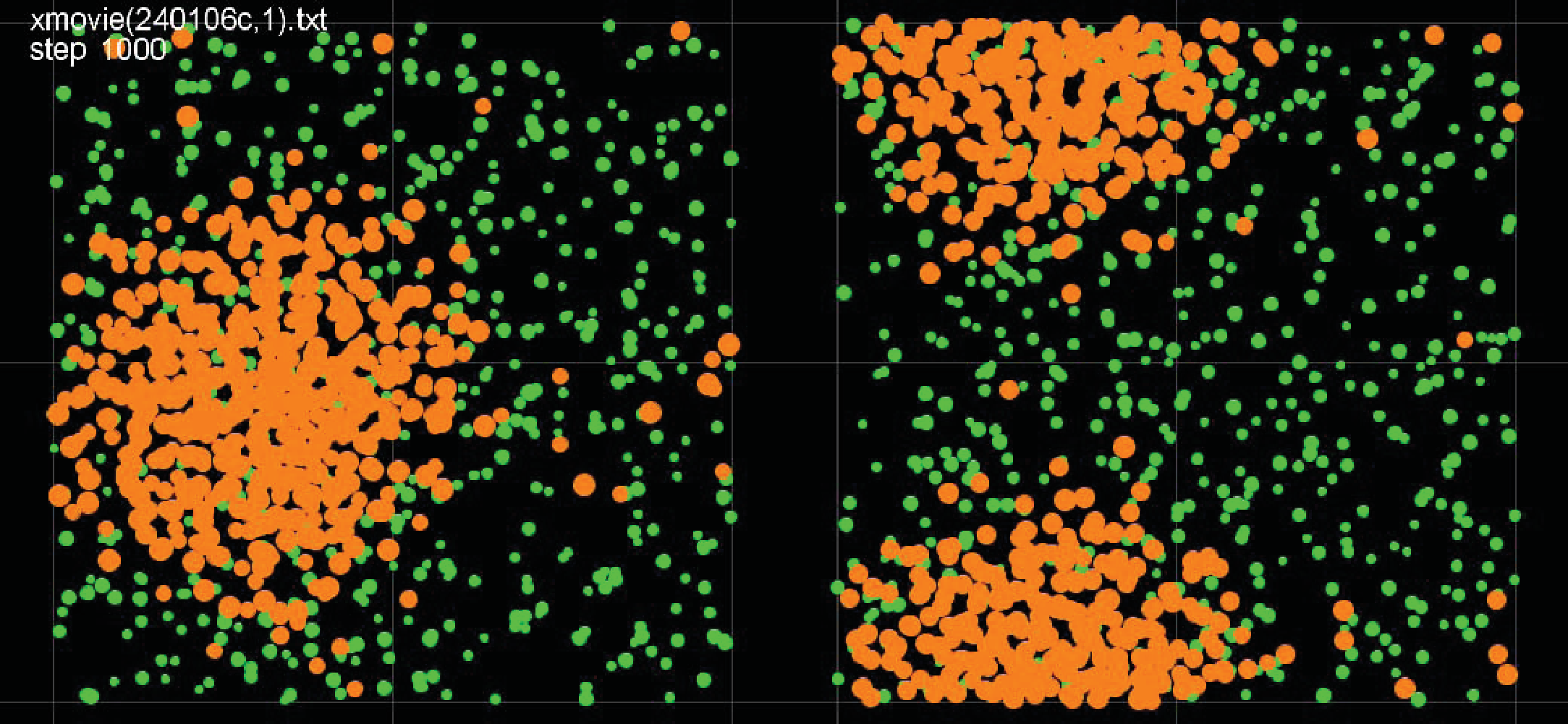}
\hspace*{0.2cm}
\caption{Distributions of fine particles 
with $r_{p2}=2\ {\rm \mu m}$ (orange) and $r_{p1}=1\ {\rm \mu m}$ (green) 
projected onto $xy$- and $xz$-planes when $n_e^{(0)}=1\cdot 10^8\ [{\rm cm^{-3}}]$
at $(\Gamma_2, \xi_2)=(199.3, 10.15)$ (left) and $(265.7, 7.61)$ (right).}
\end{figure}
\begin{figure}[h]
\vspace*{0.5cm}
\hspace*{0.2cm}
\includegraphics[width=80mm]{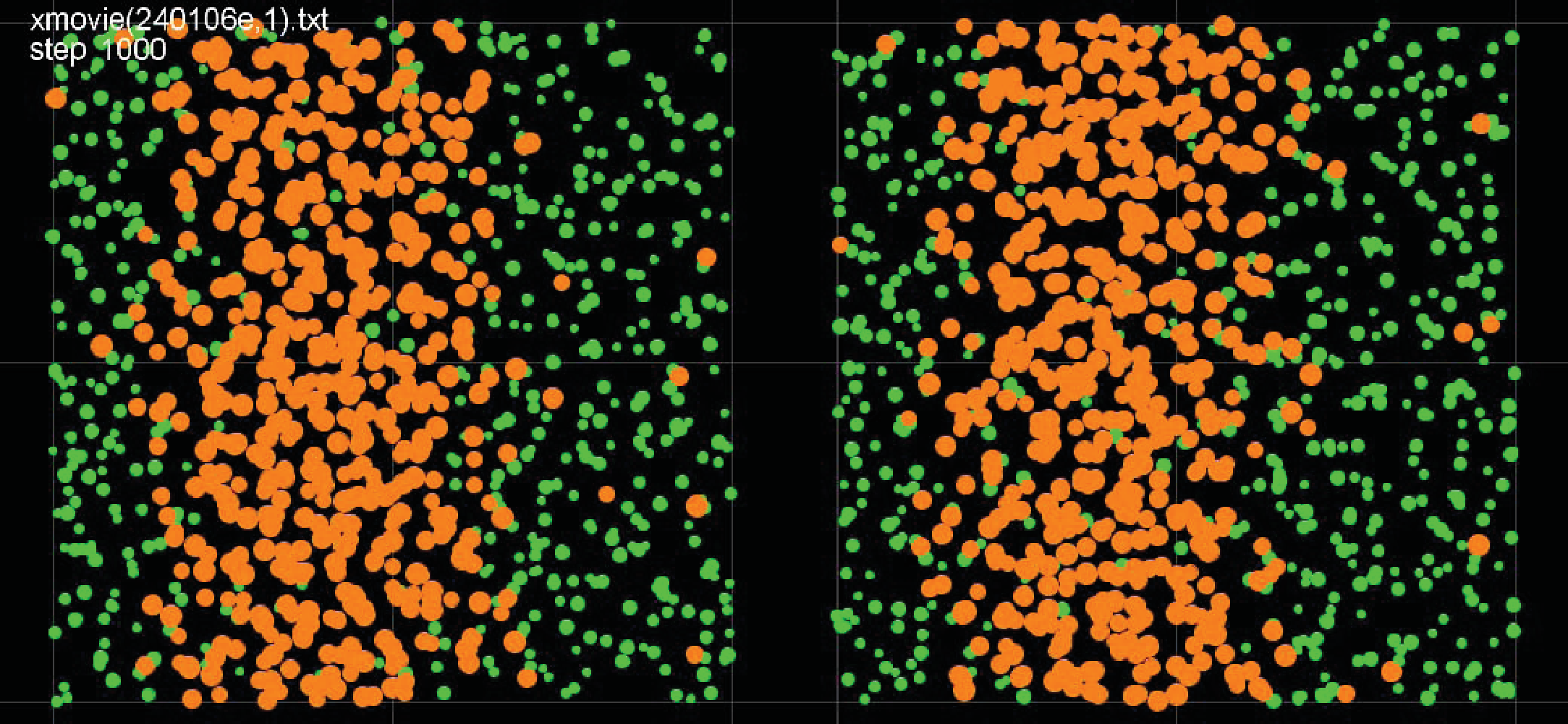}
\hspace*{1cm}
\includegraphics[width=80mm]{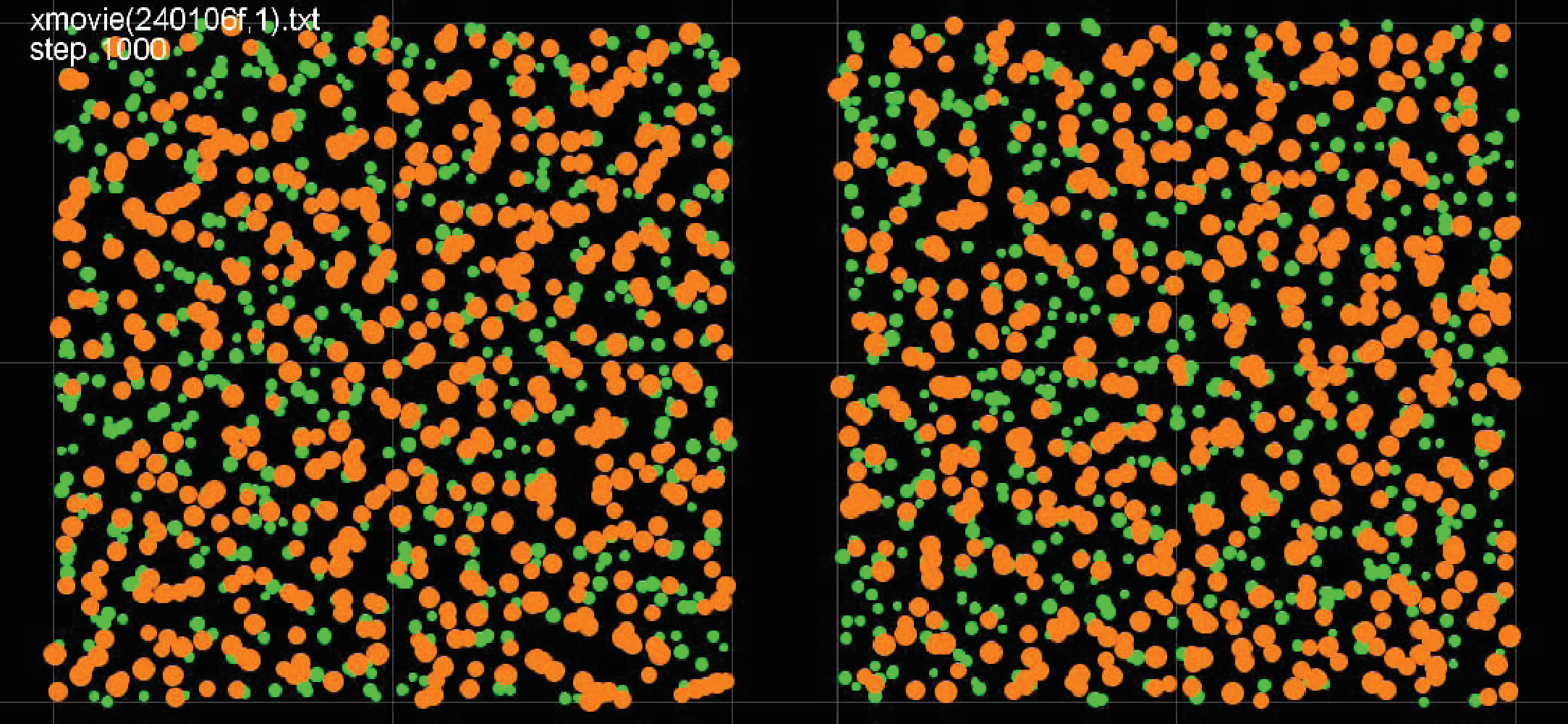}
\hspace*{0.2cm}
\vspace*{0.5cm}
\caption{The same as Fig,3 at $(\Gamma_2, \xi_2)=(398.6, 5.073)$ (left) and $(531.5, 3.805)$ (right).}
\end{figure}

The structure factors $S_{\alpha \beta}({\bm k})$ for species $\alpha$ and $\beta$
are defined by
\begin{align}
S_{\alpha \beta}({\bm k})
=\frac{<\rho_\alpha({\bm k}) \rho_\beta(-{\bm k}) >}{(N_\alpha N_\beta)^{1/2}}
-(N_\alpha N_\beta)^{1/2}\delta_{{\bm k},0},
\end{align}
where
\begin{align}
\rho_\alpha({\bm k}) =\sum_{i=1}^{N_\alpha}\exp( i{\bm k}\cdot{\bm r}_{\alpha, i}) \end{align}
and the wave number \{${\bm k}$\} corresponds to the periodicity of the system.
The pair distribution functions $g_{\alpha \beta}({\bm r})$
and the structure factors $S_{\alpha \beta}({\bm r})$ are related by
\begin{align}
n_\beta [g_{\alpha \beta}({\bm r})-1]
= \frac{1}{V}\sum_{{\bm k}}\exp( i{\bm k}\cdot{\bm r})
[S_{\alpha \beta}-\delta_{\alpha \beta}].
\end{align}
Phase separations are indicated 
by the increase in the correlations and density fluctuation amplitudes of larger particles
as shown in Figs.5 and 6.
In these figures,
the mean distance of fine particles $a_p$ is simply denoted by $a$.
\begin{figure}[h]
\hspace*{2cm}
\includegraphics[width=50mm]{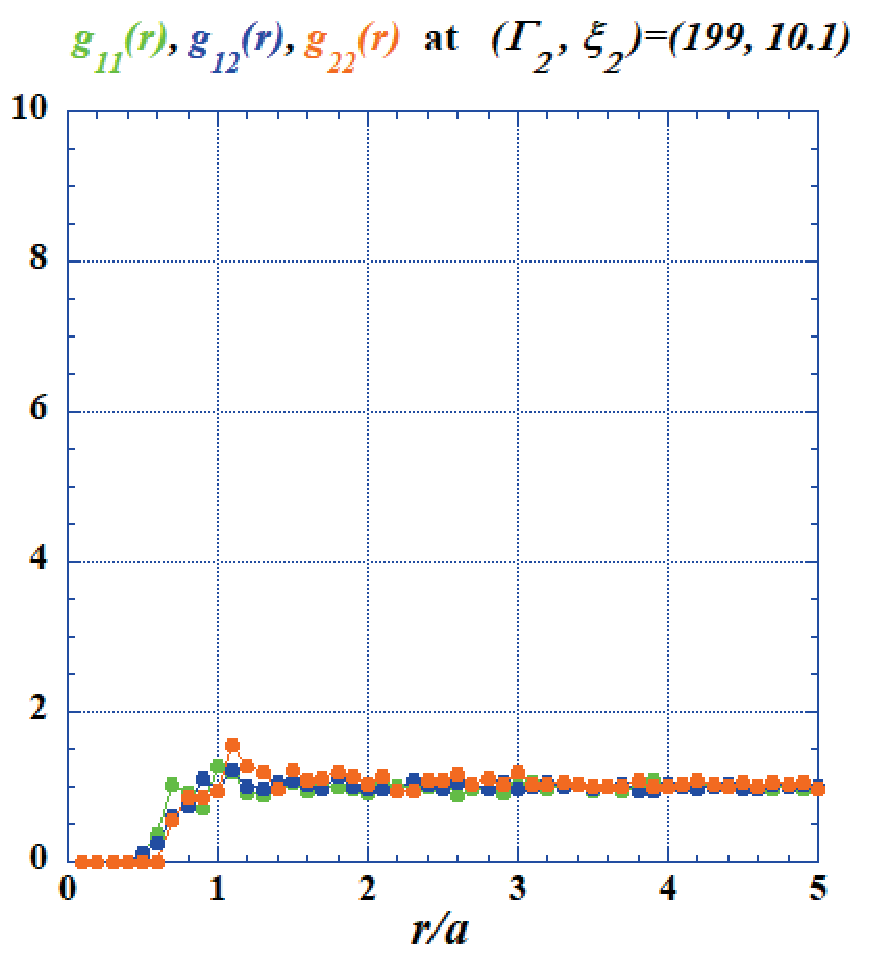}
\hspace*{2cm}
\includegraphics[width=50mm]{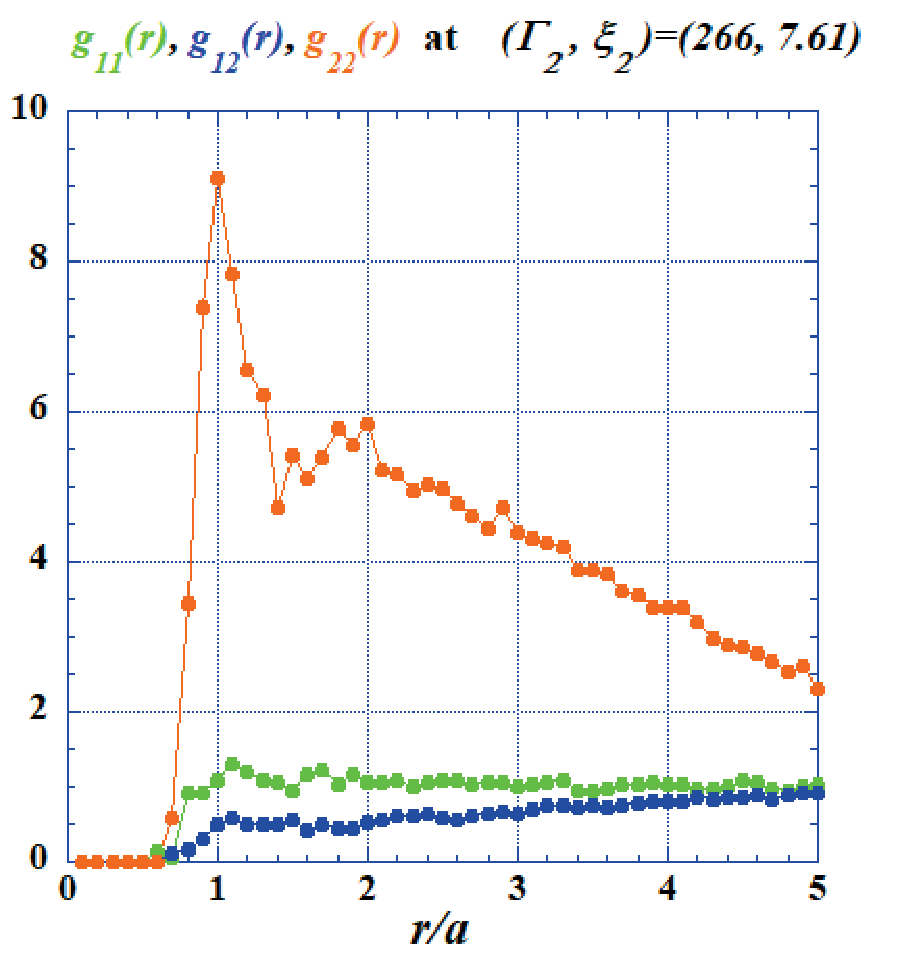}
\caption{Changes in pair distribution functions $g_{\alpha \beta}({\bm r})$
near the boundary of phase separation $(x=0.5)$.
Pair distribution functions in the domain of no separation (left)
 in comparison with those in the phase-separated domain (right).}
\end{figure}
\begin{figure}[h]
\includegraphics[width=39mm]{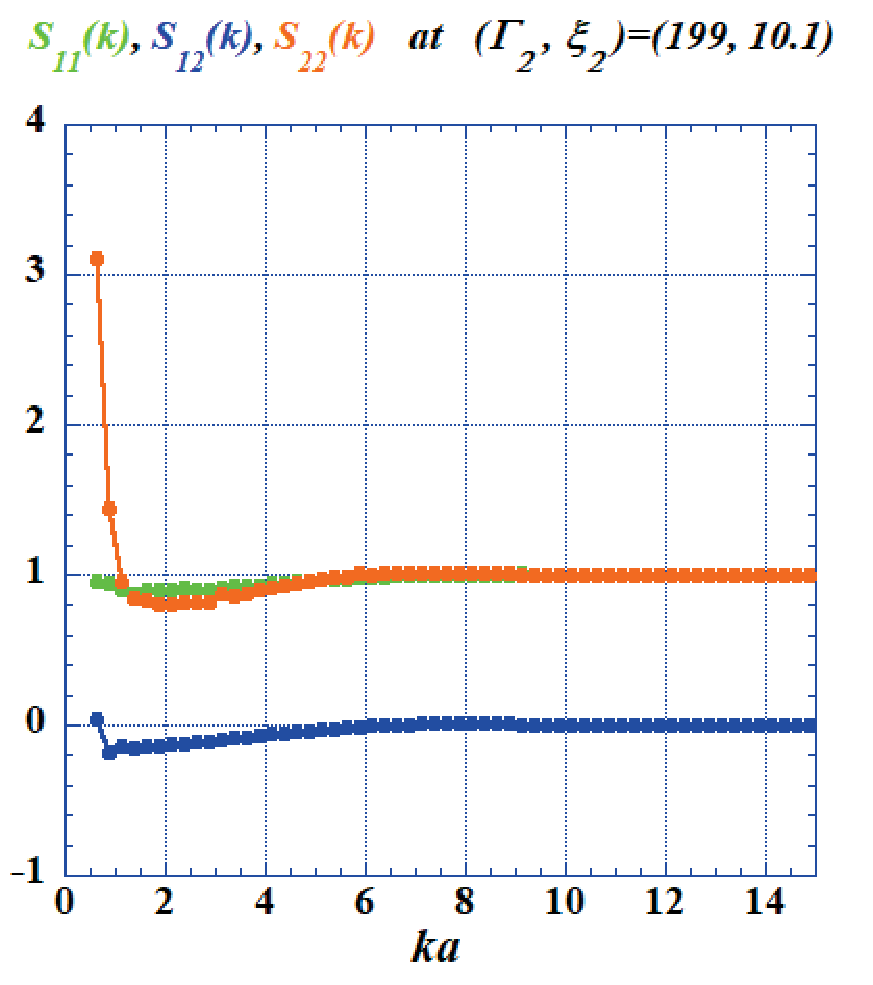}
\includegraphics[width=40mm]{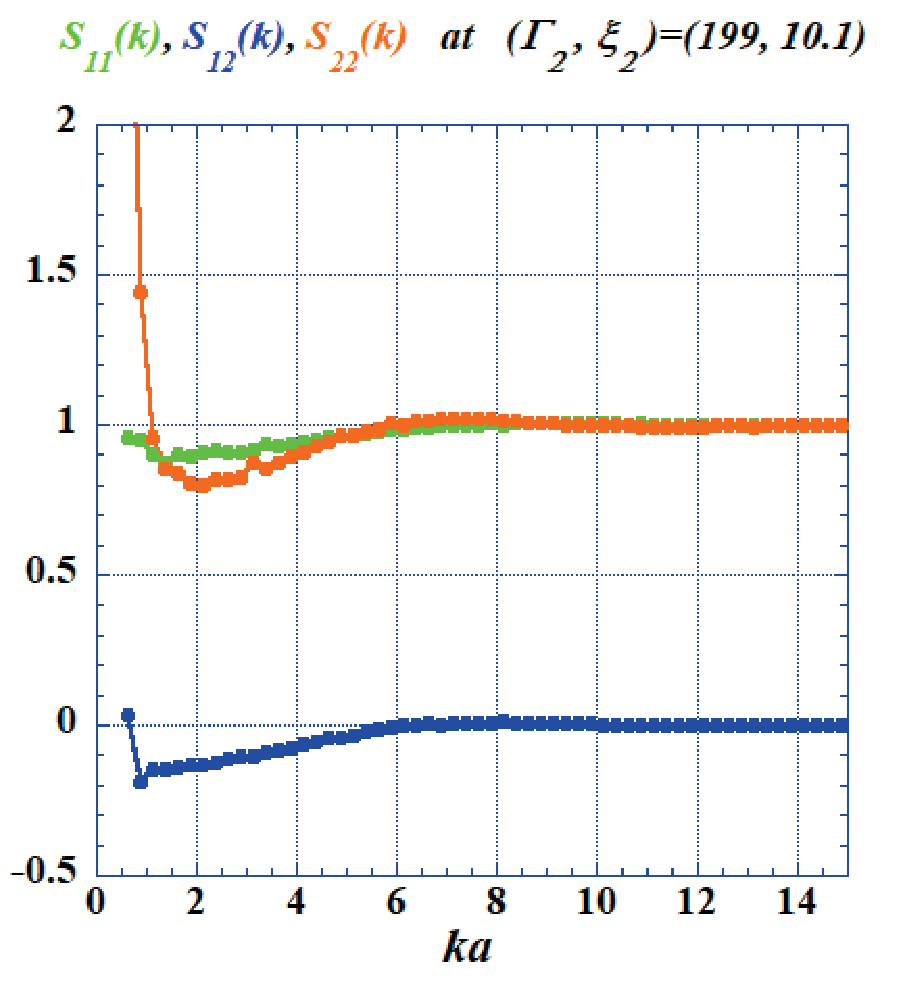}
\hspace*{1cm}
\includegraphics[width=40mm]{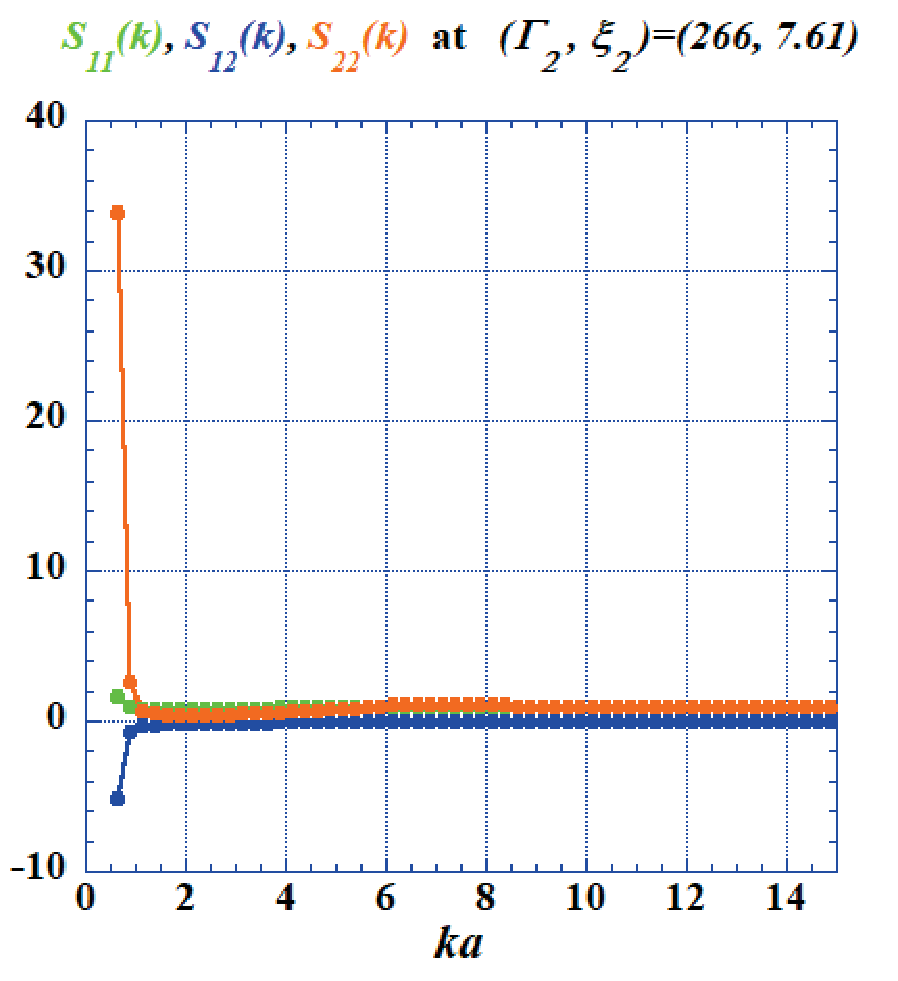}
\includegraphics[width=41mm]{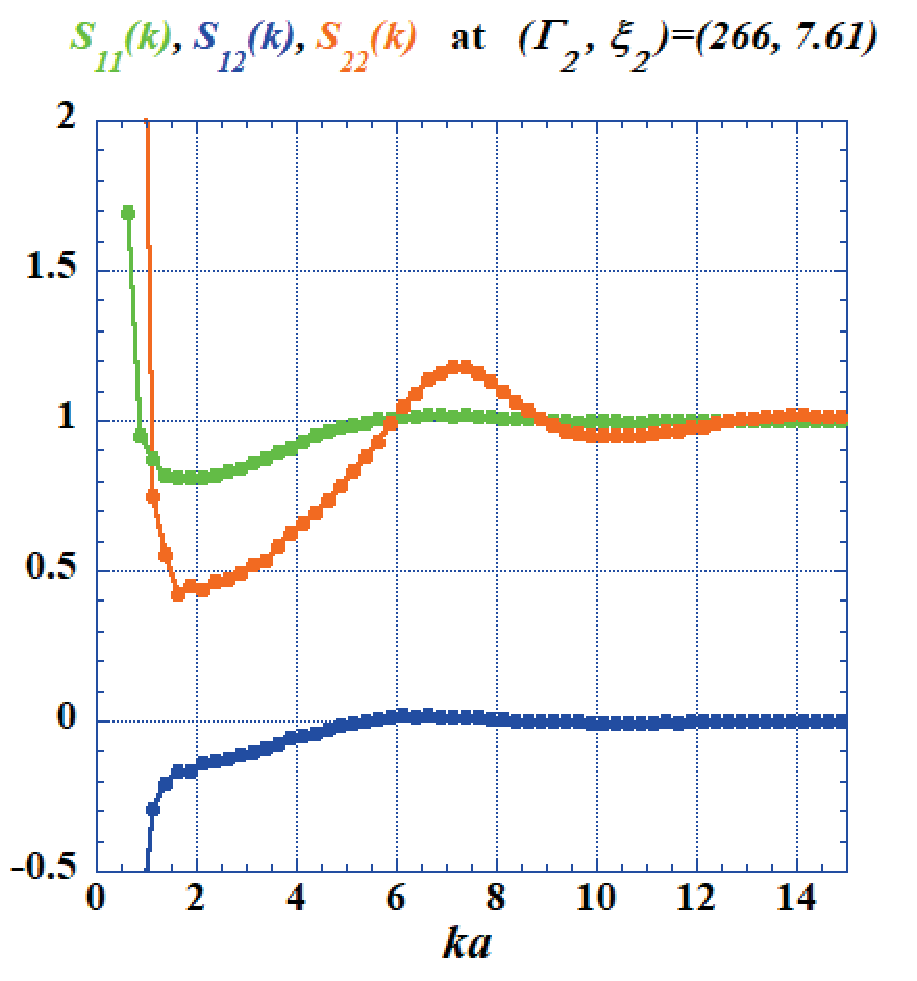}
\caption{
Changes in structure factors $S_{\alpha \beta}({\bm k})$
near the boundary of phase separation $(x=0.5)$.
Structure factors in the domain of no separation (left pair)
 in comparison with those in the phase-separated domain (right pair).}
\end{figure}

Phase diagrams for $0.25 \leq x \leq 0.75$ are shown in Fig.7:
At red points, we have separated phases and
fine particles are mixed at blue points.
We observe that, at least in this range of mixing parameter $x$,
the phase diagrams are similar to the case of one-component system
shown in Fig.2.

In relation to the above observation,
it is to be noted that, even $Q_1 e n_1 \ll e n_e = e n_i^{(0)}$, 
we have $(Q_1 e)^2 n_1 \gg e^2 n_e=e^2 n_i^{(0)}$.
The latter inequality indicates the possibility 
that the screening of fine particles of species 2 is strongly influenced 
by the existence of fine particles of species 1
which have larger Debye wave number than the ambient electron-ion plasma.
Our results, however, are not so different from the one-component case.
We may understand this fact considering the fact that, 
when strongly coupled as fine particles of species 1 in our case, 
the simple Debye screening does not work:
For example, the correlation functions in strongly coupled plasmas cannot be interpreted
in terms of Debye screening
as in the case of weak coupling. 

\begin{figure}[h]
\includegraphics[width=57mm]{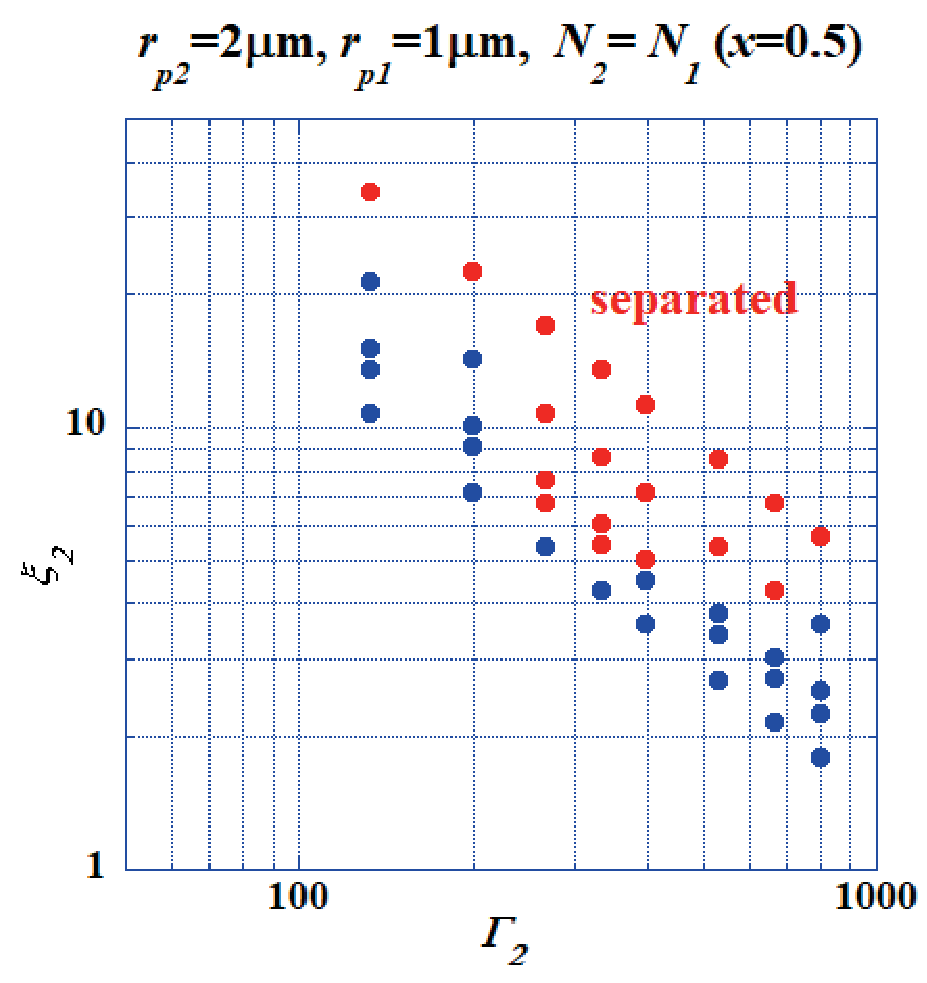}
\includegraphics[width=58mm]{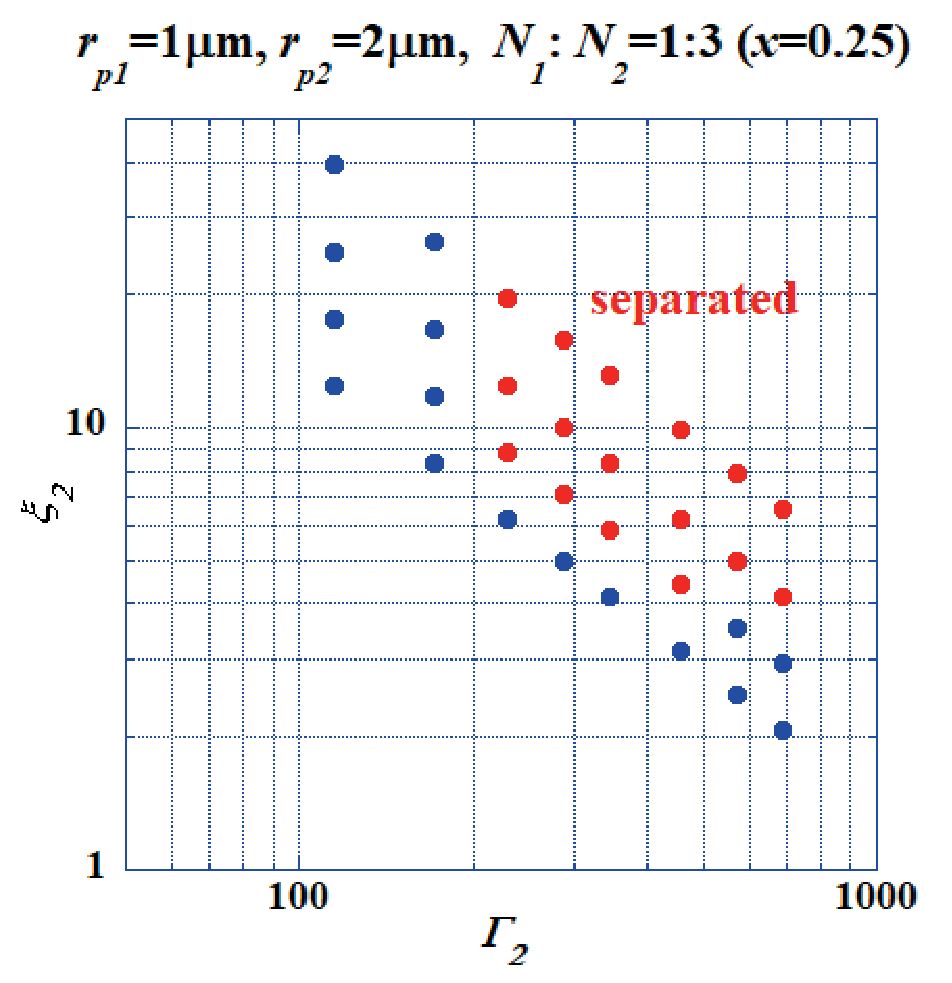}
\includegraphics[width=58mm]{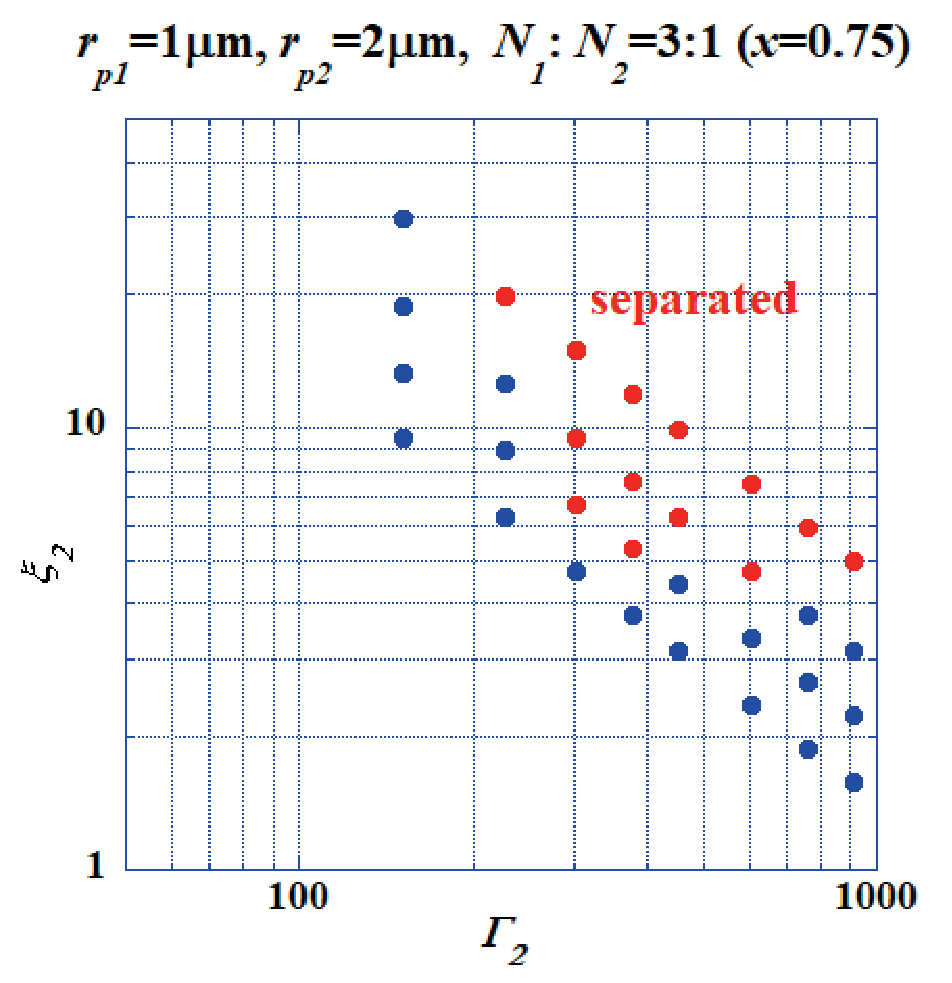}
\caption{
Phase diagrams for mixtures of fine partices 
with radius $1\ {\rm \mu m}$ and $2\ {\rm \mu m}$ at $k_BT_e=3\ {\rm eV}$.
Mixing ratio $x$ are, from left to right,  0.5, 0.25, and 0.75.}
\end{figure}

\label{sec3}
\section{Discussions}

When the ambient electron-ion plasma is uniform,
the distribution of fine particles is also uniform due to the requirement of charge neutrality
and
we have the fluid-solid transitions
with the solid phase of either the face-centered cubic or the body centered cubic
and
obtain the phase diagram of the Yukawa system\cite{RKG88etc}.

When only the electron density is uniform and the ion density can change
so as to satisfy the charge neutrality of fine particle clouds
composed of fine particles, ions, and (uniform) electrons,
we have the formation (separation or condensation) of fine particle clouds 
from the ambient electron-ion plasma.
Examples of phase diagrams obtained theoretically and by numerical simulations are shown
in Fig.1(Right) and Fig.2, respectively.

We note that,
when we have fine particle clouds,
there naturally exist only few fine particles in the ambient plasma outside of clouds
and
the latter can be regarded as the so-called void(s).
In experiments under microgravity,
voids are often observed
and
they are usually attributed to the effect of ion flow on fine particles\cite{FIKKM05}
(and also to that of thermophoretic force, 
if we have inhomogeneous temperature distribution\cite{SM02}).
Our results of formation of fine particle clouds, however,
are due to thermodynamics of fine particles (accompanying increased part of ions)
as a solute in the solvent of ambient electron-ion plasma\cite{HT20}.
We emphasize that,
in order to observe condensation of fine particle clouds as a solute,
it is necessary to somehow avoid the effect of ion flow:
Under the influence of the ion flow,
the formation of fine particle clouds might be indistinguishable from the formation of voids.
 
As for phase separation of binary mixtures,
we have spinodal decompositions due to the differences in mutual interaction potentials\cite{IZTM09}.
In our simulations,
mutual Yukawa repulsion $v_{\alpha \beta}(r)$between species $\alpha$ and $\beta$
coming from $\Phi(\{ {\bm r}_i\})$ exactly satisfy the relation
\begin{align}
v_{\alpha \beta}(r)= \left( v_{\alpha \alpha}(r) v_{\beta \beta}(r)\right)^{1/2}
\end{align}
and therefore they promote {\it neither mixing nor demixing} of two species.

In order to confirm that
the effect of asymmetry in mutual interactions can be correctly simulated,
we have tentatively modified mutual interactions between species $\alpha$ and $\beta$, $v_{\alpha \beta}$,
to $f_{\alpha \beta}v_{\alpha \beta}$
with two examples in the directions of promoting demixing,
\begin{align}
f_{11}=f_{22}=0.9,\ f_{12}=f_{21}=1.08,\ {\rm and}\ f_{11}=f_{22}=0.8,\ f_{12}=f_{21}=1.16, 
\end{align}
or with two examples in the direction of promoting mixing,
\begin{align}
f_{11}=f_{22}=1.1,\ f_{12}=f_{21}=0.92,\ {\rm and}\  f_{11}=f_{22}=1.2,\ f_{12}=f_{21}=0.84.
\end{align}
Examples of results are shown in Fig.8
where
parameters and initial conditions are identical with those in the right pair in Fig.3.
We clearly observe that
the modifications work to promote separation or mixing as expected.
At the same time,
these results indicate that
we are observing the formation of clouds of larger particles with larger mutual coupling
in the binary mixture of fine particles
and {\it not} the demixng of two species due to asymmetry of interactions.

\begin{figure}
\begin{center}
\includegraphics[width=10cm]
{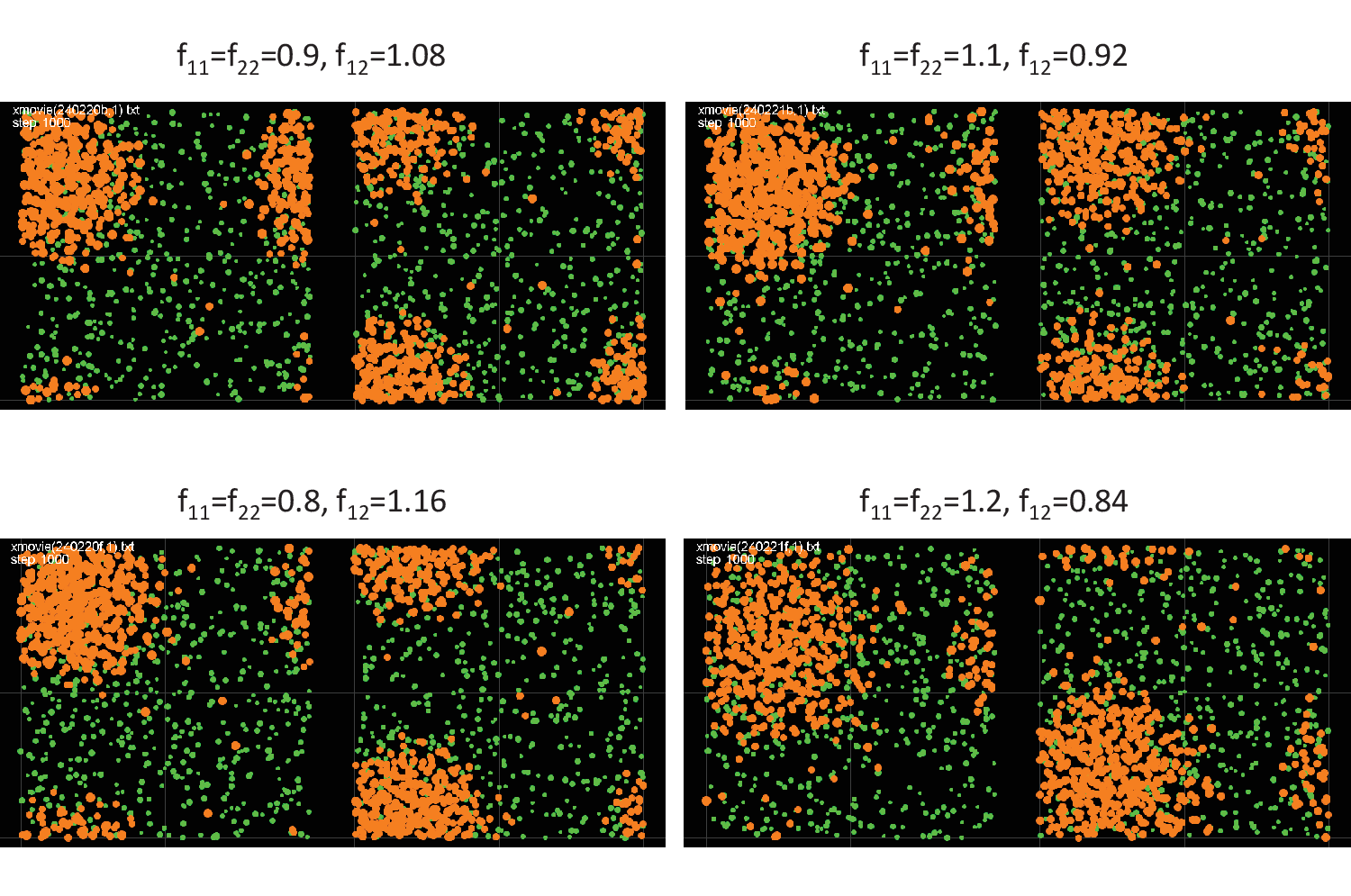}
\caption{Effect of modified mutual interactions:
Enhanced separation (demixing) in the left pairs
and
enhanced mixing in the right pairs.
Parameters and initial conditions are identical with those of the right pair in Fig.3:
$n_e^{(0)}=1\cdot 10^8\ [{\rm cm^{-3}}]$
at $(\Gamma_2, \xi_2)=(265.7, 7.61)$.}
\end{center}
\end{figure}

\label{sec4}

\section{Conclusion}
As a strongly coupled Coulomb-like system,
fine particles in classical plasmas have possibilities of phase separations.
Based on analyses on thermodynamic properties of Yukawa system,
we expect two kinds of phase separations in the domain of strong coupling:
Separations into phases with different values of densities of electrons, ions, and fine particles,
and the ones into phases with a common electron density.
The conditions for these possibilities have been clarified
by analyzing behavior of four components of the system.
The results of separations into phases with a common electron density
have been confirmed by numerical simulations.

Here we have extended numerical simulations to the case of two components.
Though the ratio of the size 
(giving approximate ratio of charges on fine particles)
is fixed at 2
and 
the mixing ratio $x$ is within the range $0.25 \leq x \leq 0.75$,
we have obtained the results indicating 
that the effect of the existence of components with smaller charges
is not so significant for the condensation of fine particles with larger charges.
We expect
these phenomena may be observed
in experiments under microgravity 
and also those on the ground as two-dimensional structures.

\section*{Acknowledgements}
The author have been inspired by participating in the series of conferences Strongly Coupled Coulomb Systems (SCCS, started as Strongly Coupled Plasma Physics at Orleans in 1977).
He would like to thank Gabor Kalman and Kenneth I. Gorden
for their contributions to the series.

\end{document}